\documentclass[10pt,a4paper]{article}

\usepackage{geometry}
\usepackage{subfigure}
\usepackage{graphicx}
\usepackage{subcaption}
\usepackage{tabularx}
\usepackage{booktabs}

\usepackage{amsmath}
\usepackage{amsfonts}
\usepackage{url}
\usepackage{indentfirst}

\usepackage[dvipsnames]{xcolor}
\usepackage[colorlinks=true, allcolors=magenta]{hyperref}

\usepackage{lineno}

\usepackage{authblk}
\author[]{Ho Fung Tsoi*}
\author[]{Kam Wai Lai}
\author[]{Dylan Rankin}
\affil[]{University of Pennsylvania, USA}

\title{\textbf{SpikePID: Real-Time Neuromorphic Cluster Counting in Future Drift Chambers}}

\date{{\normalsize*E-mail: ho.fung.tsoi@cern.ch}}

\begin{document}

\maketitle

\begin{abstract}
{\normalsize
Cluster counting ($dN/dx$) measures discrete ionization occurrences along a charged particle track traversing a gaseous medium and is currently being proposed to serve particle identification (PID) in next-generation drift chambers, as it is expected to offer higher separation power than the traditional approach based on the ionization energy loss measurement ($dE/dx$), which has intrinsically larger statistical fluctuations. However, measuring $dN/dx$ in drift chambers proposed for future colliders requires high-resolution waveforms, which result in very large data rates. Running the reconstruction at the edge would drastically reduce the bandwidth required to transmit data off the detector. In this paper, we introduce SpikePID, a neuromorphic computing approach applying spiking neural networks (SNNs) to this task due to their inherently sparse computing and processing nature. We show that an SNN with only 529 parameters improves the kaon-pion separation power by 17\% over the traditional derivative-based method and by 4\% over an MLP that has $8.6\times$ more parameters. When synthesized to an FPGA, it uses an order of magnitude fewer resources than the MLP, and has an inference latency compatible with the constraints of future collider experiments, making the SNN an excellent algorithmic candidate for real-time data readout in future drift chambers.
}
\end{abstract}

\noindent\rule{\textwidth}{1pt}
{\hypersetup{linkcolor=blue}\normalsize\tableofcontents}
\noindent\rule{\textwidth}{1pt}

\clearpage

\section{Introduction}
\label{sec:introduction}

Gaseous drift chambers have long served as a crucial sub-detector component in high-energy physics (HEP) experiments, providing essential capabilities for charged particle tracking, momentum measurement, and particle identification (PID)~\cite{BaBar:2001yhh,ABASHIAN2002117,Adinolfi:2002uk,CDF:2003xbh,Belle-II:2010dht}.
PID is expected to be critical to the performance of tasks like flavor tagging of quark jets for precision Higgs measurements~\cite{FCC:2025lpp}.
These capabilities are made possible by high spatial and timing resolutions, which enable precise measurements of ionization activities within a large volume at a low material budget.
In the context of future collider planning, such as the electron-positron phase of the Future Circular Collider (FCC-ee)~\cite{FCC:2018evy} and the Circular Electron Positron Collider (CEPC)~\cite{CEPCStudyGroup:2018ghi}, conceptual designs for next-generation drift chambers have been proposed and are being optimized.
These designs feature significantly higher precision and granularity to deliver the necessary physics performance under the unprecedented luminosities expected at these future colliders.
In particular, a primary objective in the design of future drift chambers is to improve the PID performance by migrating the underlying technique from traditional energy loss measurements ($dE/dx$) to more sensitive cluster counting measurements ($dN/dx$).
However, cluster counting raises new practical challenges because it requires exceptionally high digitization rates, typically at $O(\text{GHz})$, to be able to resolve individual ionization clusters~\cite{IDEAStudyGroup:2025gbt}.
The high rates drastically inflate the volume of raw data produced by the detector, which adds a significant additional bandwidth requirement in the real-time data readout system.
In principle, reading out the full waveforms may be possible, but this large data volume could already be avoided at the source by running on-detector reconstruction of the cluster counts.
Therefore, we expect that a feasible cluster counting algorithm should not only reconstruct the cluster counts accurately to deliver effective PID performance, but ideally also run efficiently on-detector in real time to avoid unnecessarily inflating the data readout bandwidth.

\begin{figure}[!t]
    \centering
    \includegraphics[width=\textwidth]{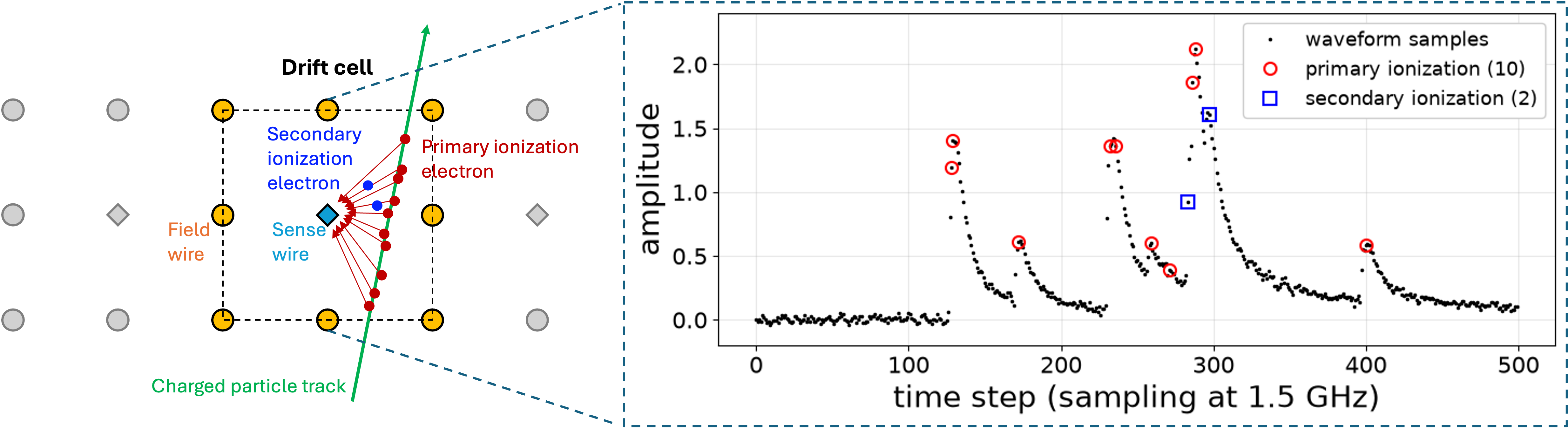}
    \caption{Example schematic diagram of a drift cell in square geometry acquiring waveform data as a charged particle traverses it. Left: cross section of a drift cell consisting of a central sense wire and eight field wires, traversed by a charged particle track, producing primary ionization electrons (red) and secondary electrons (blue) that drift towards the sense wire under an electric field. Right: an example simulated waveform digitized at 1.5 GHz, showing the truth labels of primary ionization (red) and secondary ionization (blue).}
    \label{fig:waveform}
\end{figure}

\textbf{Drift chamber basics.}
The generic working principle of a drift chamber~\cite{Rolandi:2008ujz} is illustrated in Fig.~\ref{fig:waveform}, which shows a cross-sectional view of a single drift cell acquiring waveform signals when a charged particle traverses it.
A typical drift chamber is cylindrical and consists of many densely packed small drift cells formed by long wires extending between the two end-plates.
Each drift cell is composed of a sense wire at the center as the anode and several field wires at the boundaries as the cathodes.
A charged particle traversing a cell ionizes the gas mixture contained inside, producing primary ionization electrons; these electrons, when energetic enough, can further ionize more gas molecules downstream, producing secondary ionization electrons.
Each primary electron and its associated secondary electrons form a localized ionization cluster.
Under the electric field applied between the sense wire and the surrounding field wires, these electron clusters drift towards the central sense wire, generating electrical pulses that are recorded and digitized into the time-series waveform shown.
In the waveform profile, each sharp amplitude peak corresponds to the arrival of an ionization electron at the sense wire, which triggers an avalanche ionization where the electrons are collected almost instantly; the heavy positive ions left behind then slowly drift away from the sense wire, inducing a decaying signal current that shows up as a long falling tail following each peak.

\textbf{PID based on traditional $dE/dx$.}
According to the Bethe-Bloch formula, the mean energy loss per unit length ($dE/dx$) of a charged particle traversing a medium via ionization depends primarily on its velocity through the factor $\beta\gamma=p/m$~\cite{ParticleDataGroup:2024cfk}.
Hence, particles with the same momentum but different masses have different $dE/dx$ rates, which can be exploited to identify the particles, e.g.\ kaons vs.\ pions.
In a drift chamber, this is typically measured by averaging the energy deposited within drift cells along the track~\cite{allison1980a}. 
While most ionization events involve small energy transfers, there is a non-zero probability that one transfers a significant fraction of energy, producing an energetic secondary electron that continues to ionize in cascade, so that the energy deposit follows a Landau distribution with a long tail~\cite{Rolandi:2008ujz}.
These energetic secondary ionization events can inflate the mean $dE/dx$ measurement, degrading its resolution.
Methods have been developed to mitigate this effect, such as the truncated mean calculations described in Ref.~\cite{allison1980a}, where a fixed fraction of the highest-energy samples is discarded to suppress contributions from the tail, but the fluctuations still remain in the retained samples.
This limitation has prompted the development of alternative techniques such as cluster counting ($dN/dx$), which has a fundamentally cleaner statistical profile.

\textbf{PID based on cluster counting $dN/dx$.}
In contrast to $dE/dx$, cluster counting ($dN/dx$) measures the number of primary ionization clusters per unit length~\cite{Rolandi:2008ujz,4329616}.
Since it counts discrete ionization occurrences instead of summing over a continuous spread of energy deposits, the method is agnostic to the fluctuations in energy deposits per cluster, so its resolution is insensitive to the Landau tail.
Moreover, the count follows Poisson statistics since these discrete ionization events occur independently along the track, and the $dN/dx$ measurement is expected to give a PID separation power about a factor of two better than the $dE/dx$ method~\cite{FCC:2018evy,IDEAStudyGroup:2025gbt}.
The traditional approach to cluster counting uses derivatives to locate peaks in a waveform and then classifies those peaks into primary ionization via thresholds~\cite{Zhao:2024ely,Yilmaz:2025kqz,Tian:2024yfz,Elmetenawee:2025ioq}.
In this work we take this derivative-based method as a baseline and use machine learning (ML) methods to further improve the performance.

\textbf{Real-time readout constraints.}
Since most next-generation drift chambers are designed to enable $dN/dx$ for better PID performance, the GHz-level sampling rate required to resolve neighboring clusters in each waveform poses significant challenges to the design of the data readout systems.
To quantify the realistic constraints expected in future collider experiments, we take the IDEA (Innovative Detector for Electron-positron Accelerator) drift chamber~\cite{IDEAStudyGroup:2025gbt} as an example.
According to the readout estimates for the IDEA drift chamber at FCC-ee in the conceptual design report~\cite{FCC:2018evy}, a hadronic $Z$ decay with 20 tracks, each hitting 130 drift cells, produces about 4 MB of digitized waveform data, requiring a bandwidth of around 400 GB/s at the $Z$ pole.
Adding other contributions from $\gamma\gamma\rightarrow\text{hadrons}$, $e^{+}e^{-}$ pair background, and noise, the total raw data rate produced at the drift chamber amounts to around 1 TB/s.
This poses a significant challenge if the full waveforms are read out and stored for offline processing.
Instead, if a cluster counting algorithm runs on-detector to perform the reconstruction in real time and transmits only the minimal information, the data to be read out can be significantly reduced.
For example, the bandwidth for hadronic $Z$ decays could already be reduced from 400 GB/s to 25 GB/s by transmitting only the timing and amplitude of the identified ionization peaks (a 16$\times$ reduction), while transmitting only one cluster count per cell would reduce the rate even further.
In order to realize this reduction, the cluster counting algorithm must be compact enough to fit on edge electronics such as FPGAs or ASICs.

\textbf{Neuromorphic computing.}
While there has been significant work in HEP and beyond to develop and deploy ML algorithms on edge devices, comparatively little focus has been given to neuromorphic computing.
Neuromorphic computing is an architectural framework of software-hardware co-design that closely follows biological neural processing~\cite{58356,indiveri2011neuromorphic,merolla2014million,schuman2017survey,roy2019towards}.
In the human brain, a neuron accumulates incoming signals and fires a discrete electrical pulse only when its membrane potential is sufficiently charged up.
Hence, neurons remain idle most of the time instead of being activated continuously, allowing the entire biological system to operate with very low power consumption to optimize energy efficiency.
This mechanism can be realized computationally via spiking neural networks (SNNs)~\cite{maass1997networks,gerstner2002spiking,burkitt2006review,tavanaei2019deep}, which use spiking neurons that integrate information over time and fire binary spikes only once a set threshold is crossed.
In contrast to conventional deep neural networks, where every neuron computes and produces outputs in every forward pass, SNNs are functionally sparse in nature and can run with very low power consumption on hardware.
When implemented on custom electronics such as FPGAs with highly parallel processing capabilities, SNNs could potentially be an excellent architecture for resource-constrained systems with high-frequency inputs, such as real-time drift chamber data readout.

This work is motivated by the following problem:
\begin{itemize}
    \item What would be a feasible cluster counting algorithm for next-generation drift chambers, which can provide better PID performance than existing methods and run on custom electronics such as FPGAs to meet the readout constraints expected in future collider experiments?
\end{itemize}
We introduce SpikePID, a neuromorphic approach to address this problem.
Our contributions are summarized as follows:
\begin{itemize}
    \item Proposed a novel approach of using SNNs to process waveform data from drift chambers and showed that their PID performance exceeds that of the traditional derivative-based method and an ML baseline using a multilayer perceptron (MLP).
    \item Developed a high-level synthesis (HLS) implementation of \textit{streaming SNNs} by extending the existing SNN support in hls4ml, so that samples are processed sequentially upon arrival without buffering the full waveform.
    \item Demonstrated the feasibility of on-detector deployment with FPGA synthesis results showing that the SNNs use an order of magnitude fewer resources compared to the MLP baseline.
\end{itemize}

The rest of the paper is structured as follows: Sec.~\ref{sec:related} discusses relevant works we took inspiration from, Sec.~\ref{sec:method} introduces SNNs and the HLS implementation, Sec.~\ref{sec:setup} presents the dataset and the experimental setup, Sec.~\ref{sec:results} presents the model performance and the FPGA synthesis results, Sec.~\ref{sec:discussion} discusses the impacts of the study and future directions, and Sec.~\ref{sec:summary} summarizes the work.

\section{Related Work}
\label{sec:related}

We first discuss the existing algorithms for cluster counting and their limitations in terms of PID performance and deployment in drift chambers, and then some SNN applications in other HEP areas that motivated this work.

\textbf{Existing cluster counting algorithms.}
A simple derivative-based method~\cite{Yilmaz:2025kqz,Tian:2024yfz} first computes derivatives over the waveform samples to locate sharply rising peaks, and then classifies these peak candidates into primary ionization clusters using selections tuned to veto secondary ionization.
Ref.~\cite{Elmetenawee:2025ioq} validated the improved performance of cluster counting over the $dE/dx$ method on test beam data.
However, simple derivative-based methods yield only modest accuracy because they are prone to falsely capturing amplified noise spikes, and the peak classification relies on fixed thresholds that are too rigid to adapt to dynamical drifting environments.
ML methods have been proposed to advance the cluster counting technique.
Ref.~\cite{Tian:2024yfz} improves the performance of the same two-step setup by training a recurrent neural network for peak finding and a graph neural network for clustering.
However, these models are far too big to fit the real-time constraints described above.
Without separate steps, Ref.~\cite{Zhao:2025gkk} uses a graph neural network with attention-based aggregation to directly classify each hit from point cloud data in a time projection chamber and derive the cluster counts.
Taking into account the real-time constraints, Ref.~\cite{Yilmaz:2025kqz} constructs a small MLP and synthesizes it into an FPGA to show manageable resource estimates.

\textbf{SNN applications in HEP and FPGAs.}
Recently, SNNs have been explored for trigger applications in collider experiments due to their potential to fit the tight computational constraints.
Ref.~\cite{10.1145/3589737.3605976} studies on-sensor SNNs in the silicon pixel system that filter out tracks with low transverse momentum to reduce the total data read out from the detector.
Ref.~\cite{Dillon:2025yqe} studies SNNs to perform anomaly detection tasks for potential low-latency trigger applications.
To realize real-time inference applications, work has been done to support compiling trained SNNs into HLS code for FPGAs.
S2N2~\cite{10.1145/3431920.3439283}, built upon the FINN library~\cite{10.1145/3020078.3021744} for accelerating binarized neural networks on FPGAs, was developed to support efficient implementation of SNN architectures on FPGAs.
Ref.~\cite{dillon2026spiking} also adds SNN support to the hls4ml library~\cite{10.1145/3801979,Duarte:2018ite}, targeting dense SNN architectures.
Our HLS implementation builds upon Ref.~\cite{dillon2026spiking} and extends it to support sample-streaming inference, as described in Sec.~\ref{sec:method_hls}.

\section{Method}
\label{sec:method}

We introduce the basics of spiking neurons and our streaming SNN architectures for the cluster counting task in Sec.~\ref{sec:method_snn}, and the HLS implementation in Sec.~\ref{sec:method_hls}.

\subsection{Spiking Neural Networks}
\label{sec:method_snn}

Like their traditional counterparts in conventional neural networks, spiking neurons are the building blocks of SNNs~\cite{maass1997networks,gerstner2002spiking,burkitt2006review,tavanaei2019deep}.
Their design is inspired by the event-driven signal processing of biological neural systems, where neurons maintain a membrane potential and emit spiking electrical pulses to convey information to other neurons once that potential is charged above a threshold by incoming spikes.
A spiking neuron is a stateful unit that holds an internal state value called the membrane potential, which evolves over discrete time steps.
At each time step, the neuron starts from the residual membrane potential inherited from the previous time steps, weighs the incoming signals, and adds them to the current state value.
If this updated potential reaches a pre-specified threshold, the neuron fires a binary spike as an output signal to downstream neurons and resets its potential for the next time step; otherwise, no output is produced for the current time step.
Because output spikes are gated by a threshold, instead of being produced continuously at every forward pass as in conventional neural networks, SNNs can operate with far lower power consumption on hardware with event-based processing, as most neurons remain idle and fire only when necessary, similar to how the human brain optimizes energy efficiency.
The exact mechanism of how a spiking neuron integrates incoming signals, updates its potential, and fires spikes depends on the specific spiking model.

\begin{figure}[!t]
    \centering
    \includegraphics[width=0.7\textwidth]{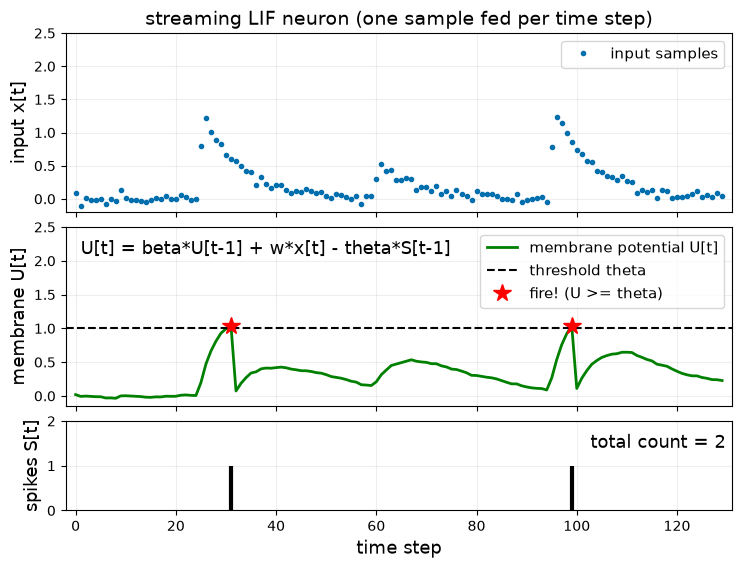}
    \caption{Response of a single LIF neuron to a toy waveform with one sample fed per time step. Top: the waveform input. Middle: the evolution of the membrane potential. Bottom: the output spikes. The LIF neuron uses $\beta=0.9$, $\theta=1$, and $w=0.24$.}
    \label{fig:lif}
\end{figure}

We consider a common spiking neuron model called \textbf{leaky integrate-and-fire (LIF)}~\cite{gerstner2002spiking,burkitt2006review,Gerstner_Kistler_Naud_Paninski_2014,8891809}.
The evolution of an LIF neuron is governed by:
\begin{equation}
    \label{eq:lif}
    U[t]=\beta\, U[t-1] +w\,x[t] -\theta\, S[t-1],\qquad S[t]=
    \begin{cases}
        1,&U[t]\geq\theta \\
        0,&U[t]<\theta
    \end{cases}.
\end{equation}
Here, $U[t]$ is the membrane potential at integer time step $t$, which consists of three terms: (1) its state from the previous time step scaled by a leak factor $0<\beta<1$ so that the history is retained while decaying, (2) the integration of the current input $x[t]$ scaled by the weight $w$, and (3) the subtraction of the threshold $\theta$ if a spike was fired at the previous time step.
Together, an LIF neuron accumulates incoming signals to charge up its membrane potential, which decays over time so that its internal state correlates more strongly with recent inputs; it fires a binary output spike once the potential is charged up to the threshold, which triggers a soft reset by subtraction.
Fig.~\ref{fig:lif} illustrates how an LIF neuron responds to a toy waveform fed sequentially, one sample per time step.

In our work, one sample of the waveform is fed per time step, which matches the real-world scenario in a drift cell where the sensor data is digitized at a fixed frequency and the samples arrive sequentially one at a time.
A streaming SNN can therefore process each sample upon arrival, without the extra buffering needed by an MLP that takes the full waveform as a flat input and has to stay idle until all samples have arrived before beginning the forward pass.
We build two variants of SNN architectures called $\text{SNN}_\text{vanilla}$ and $\text{SNN}_\text{conv}$, where the former uses a dense layer to embed one sample per time step, and the latter uses a 1D convolutional layer that is capable of embedding several neighboring samples simultaneously per time step.
In each variant, we consider two model sizes: ``lite'' and ``base'', having one and two hidden layers, respectively.
In total, we have four models: $\text{SNN}_\text{vanilla}^{\text{\{lite,base\}}}$ and $\text{SNN}_\text{conv}^{\text{\{lite,base\}}}$.

\begin{figure}[!t]
    \centering
    \includegraphics[width=0.7\textwidth]{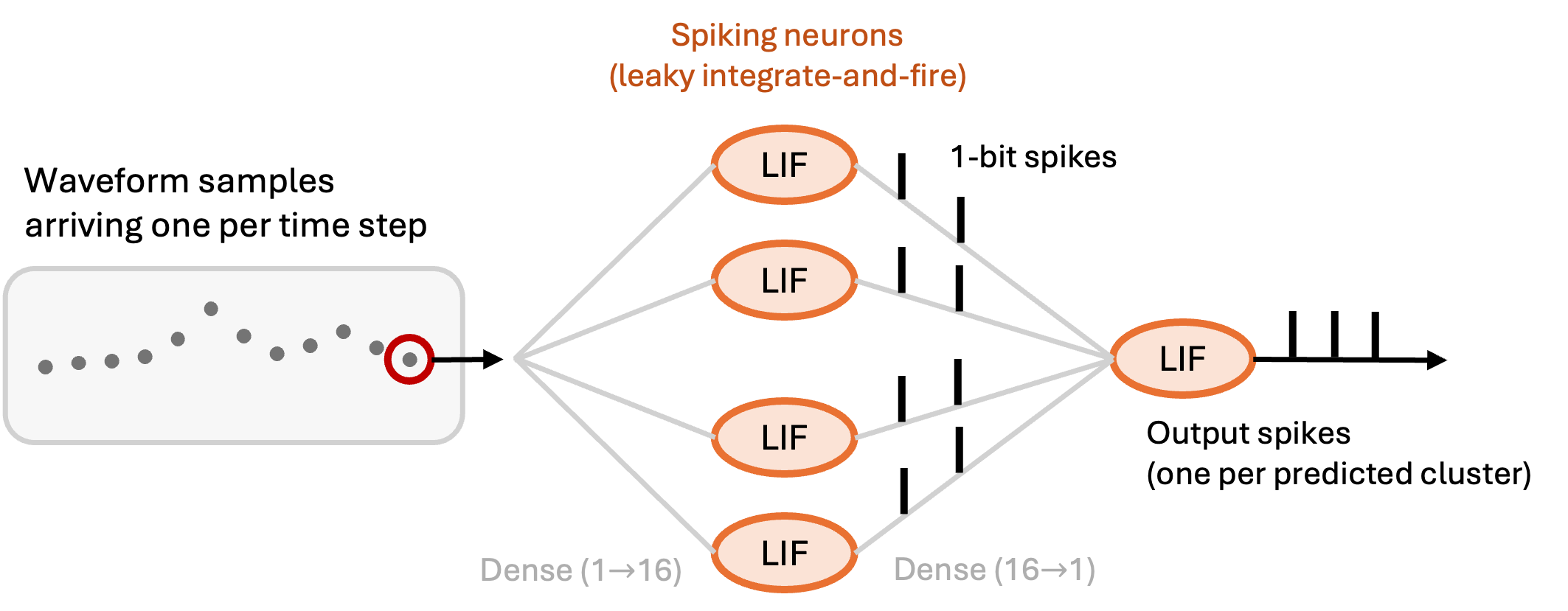}
    \caption{Model architecture of $\text{SNN}_\text{vanilla}^{\text{lite}}$, which has 17 LIF membrane states and 49 parameters.}
    \label{fig:model_vanilla}
\end{figure}

\begin{figure}[!t]
    \centering
    \includegraphics[width=1\textwidth]{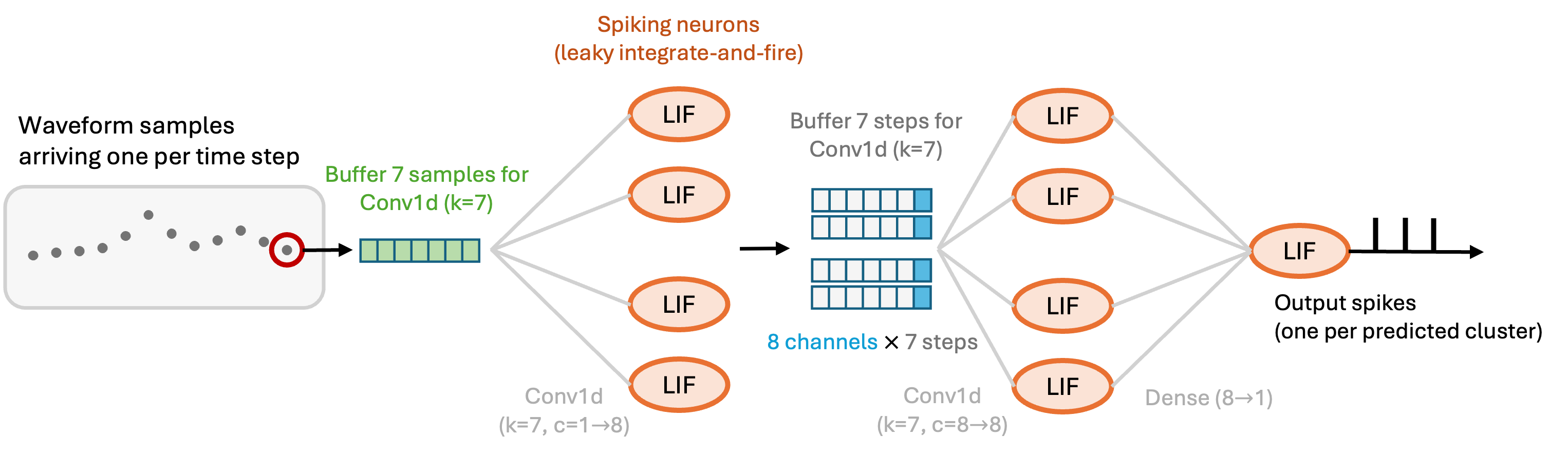}
    \caption{Model architecture of $\text{SNN}_\text{conv}^{\text{base}}$, which has 17 LIF membrane states and 529 parameters.}
    \label{fig:model_conv}
\end{figure}

Fig.~\ref{fig:model_vanilla} shows the $\text{SNN}_\text{vanilla}^{\text{lite}}$ model.
Each one-dimensional sample is embedded into 16 dimensions by a linear layer, followed by one LIF neuron per dimension.
The 16 neurons produce 16 output spike trains, which are then aggregated by a linear layer into a one-dimensional output with another LIF neuron.
The model outputs binary spikes as new samples arrive, and the model is trained so that the number of output spikes matches the truth number of primary clusters in a waveform.
The multiple channels in the hidden layers are meant to capture the complex local waveform shapes and noise, and a single LIF neuron as in Fig.~\ref{fig:lif} is not expressive enough for this and can only capture simple high peaks in a waveform.
$\text{SNN}_\text{vanilla}^{\text{base}}$ shares the same architecture but has one additional hidden layer.
Fig.~\ref{fig:model_conv} shows the $\text{SNN}_\text{conv}^{\text{base}}$ model.
The input embedding uses a causal 1D convolutional layer with 8 filters, kernel size 7, and stride 1.
This enables the model to embed seven neighboring samples per time step to better encode the local shape information and separate clusters from noise.
This produces eight LIF channels in the first hidden layer, and the second hidden layer processes them with another 1D convolutional layer with the same number of filters and kernel size, processing seven consecutive time steps per window to produce eight output channels with LIF neurons, which are finally aggregated by a linear layer into a single output as in the vanilla models.
$\text{SNN}_\text{conv}^{\text{lite}}$ shares the same architecture but has only one hidden layer.
A sweep on the model size was run to choose these SNN models: 4 to 32 hidden units and 1 to 3 hidden layers for the vanilla models, 2 to 16 channels and 1 to 3 hidden layers for the convolutional models.
The lite variants are the smallest models on the performance plateau, and the base variants have one more hidden layer.

For SNN training we use snnTorch~\cite{eshraghian2021training}, a library built on PyTorch~\cite{paszke2019pytorch} to support gradient-based learning for models with spiking neurons.
All SNN models are trained in floating-point precision and then quantized for FPGA implementation.

\subsection{HLS Implementation}
\label{sec:method_hls}

For the HLS implementation, we build upon the existing SNN support~\cite{dillon2026spiking} in hls4ml~\cite{10.1145/3801979,Duarte:2018ite}.
Before this work, hls4ml provided basic support for converting LIF neurons from snnTorch to an HLS kernel.
This kernel is compatible only with dense layers, and each function call runs a single time step forward to update the membrane potentials declared as static state within a fixed interval of time steps for a given input.
However, when a streaming SNN is designed to process inputs sequentially, the consecutive time steps cannot be pipelined under this construction, which inflates the total latency, especially when the input sequence involves many time steps.

In this work, we extend the SNN support in hls4ml in order to build the streaming SNNs used in the experiments.
The main contributions are summarized below.
\begin{itemize}
    \item \textbf{Delayed reset in membrane potential.}
    The existing firmware resets the potential in the same time step in which a spike is fired, which is one time step earlier than the default in snnTorch. We move the reset in HLS from the time step of the spike to the next one, matching snnTorch. This offset is negligible over a few time steps, but it compounds over hundreds of time steps and inflates the prediction errors.
    
    \item \textbf{Single streaming block for pipelining time steps.}
    With the membrane potential declared as static state and the top function design executing one time step only, consecutive time steps cannot be pipelined, so the total latency equals the latency per time step multiplied by the number of time steps. Therefore, we build a new template that generates a single HLS block that consumes one sample per clock cycle, wraps all time steps into an internal loop, and keeps the membrane potentials as local registers. The loop over time steps can then be pipelined, so a new sample is accepted every clock cycle without being blocked by earlier samples that are still being processed. As a result, the total latency and initiation interval (II) are greatly reduced.
    
    \item \textbf{Multiply-free spike arithmetic.}
    Spikes are stored as 1-bit values, and every spike-weight product in a layer following a spiking neuron layer is implemented as a conditional addition of the weight, so these operations become multiply-free by default and save FPGA resources.
    
    \item \textbf{Causal 1D convolution compatibility.}
    We also add support for causal 1D convolutional layers to be compatible with the streaming spiking neuron layers, so the SNN models can aggregate consecutive time steps for richer representations. The implementation keeps only a line buffer of the necessary consecutive time steps (the kernel size) instead of the full waveform.
    
\end{itemize}

\section{Experimental Setup}
\label{sec:setup}

\subsection{Dataset}
\label{sec:setup_dataset}

We use the public drift chamber dataset~\cite{da6731eab4ba495ea2dd3becc253c747} introduced in Ref.~\cite{Tian:2024yfz}, created for developing cluster counting algorithms at future colliders.
It contains waveform data generated at drift chamber cells with Garfield++~\cite{garfieldpp,Pfeiffer:2018yam}, a standard package for simulating particle interactions in gaseous detectors.
The simulation is based on the design of the CEPC drift chamber~\cite{CEPCStudyGroup:2025kmw}, where each drift cell is formed by a central sense wire and eight field wires arranged in an approximate square geometry in cross section with 18 mm side length (similar to the drawn example in Fig.~\ref{fig:waveform}).
In the simulation, the drift chamber is filled with a low-density gas mixture of 90\% He and 10\% $i\text{C}_4\text{H}_{10}$, which gives a high resolution in primary ionization clusters at a low material budget.
The waveform is sampled at 1.5 GHz (every $^2/_3$ ns), which is required to resolve peaks of neighboring primary ionization clusters at the given drift timing.
In the dataset, there are 500k pion waveforms with momenta ranging from 0.2 to 20 GeV for training, and 100k waveforms each from pions and kaons at each of seven momentum values ranging from 5 to 20 GeV in steps of 2.5 GeV for testing.
Each waveform consists of 3000 samples from a drift cell, corresponding to a time window of 2000 ns, and each sample is a numerical value representing the digitized signal output, as shown in Fig.~\ref{fig:waveform}.
Each sample has a 3-class truth label (primary ionization, secondary ionization, and background), and the number of truth primary ionization labels of a waveform is its truth cluster count.

\subsection{Models}
\label{sec:setup_models}

We use the four SNN models described in Sec.~\ref{sec:method_snn}: $\text{SNN}_\text{vanilla}^{\text{\{lite,base\}}}$ and $\text{SNN}_\text{conv}^{\text{\{lite,base\}}}$.
We consider two additional methods for comparison: the derivative-based method and an MLP.

The derivative-based method serves as the non-ML baseline that sets the minimum performance floor to validate the ML models.
We adapt the two-step scheme from Ref.~\cite{Yilmaz:2025kqz}.
In the peak-finding step, the second derivative is computed on a 5-sample moving average and a threshold is applied to locate steeply rising peak candidates, which likely originate from ionization clusters.
In the clustering step, a small window is used to scan the waveform from the first sample to the last, and peak candidates falling within the window are merged into a single primary cluster candidate.
To account for the widening of the clusters caused by diffusion, the window size increases across the waveform.
Both the second-derivative threshold and the window size are tuned to maximize the counting accuracy on 5000 waveforms in the training set.

For the ML baseline, we adapt the MLP model from Ref.~\cite{Yilmaz:2025kqz}, which is a 4-layer MLP with 8, 32, and 8 hidden dimensions with ReLU activations, followed by a single regression output.
This model size was chosen to be small enough to fit on the FPGA with computations fully parallelized while being able to perform sensible cluster counting.
The sizes of all ML models are compared in Tab.~\ref{tab:separation}.

\subsection{Model Training and FPGA Synthesis}
\label{sec:setup_hls}

Following Ref.~\cite{Yilmaz:2025kqz}, we preprocess the dataset by truncating the 3000-sample waveform to keep the first 500 samples per waveform so that the total drift time is consistent with the smaller cell size of the IDEA drift chamber at FCC-ee.
We use 400k pion waveforms for training and 50k for validation.
The inputs are standardized with the mean and standard deviation derived from the training set.
For the SNN models, we fix the LIF parameters $\beta=0.9$ and $\theta=1$.
The SNN models are configured at floating-point precision and are trained with snnTorch~\cite{eshraghian2021training} with the mean square error (MSE) loss between the number of output spikes and the truth number of primary clusters, where the gradient is backpropagated through the 500 time steps with a surrogate gradient for the spike step function~\cite{8891809}.
We use the Adam~\cite{kingma2014adam} optimizer to train with a learning rate of $10^{-3}$ and a batch size of 256 for 125 epochs, where early stopping is applied using the validation loss.
The MLP model is trained using HGQ~\cite{10.1145/3748173.3779200} for high-granularity quantization-aware training (QAT), with the MSE loss between the regression output and the truth number of primary clusters, and Adam is used to train with a learning rate of $6\times 10^{-4}$ and a batch size of 256 for up to 75 epochs with early stopping.

The MLP is then converted into HLS using hls4ml, and the SNN models are converted with the streaming SNN implementation we developed and described in Sec.~\ref{sec:method_hls}.
The SNN models are implemented as streaming blocks where one sample is consumed per clock cycle, and the MLP is implemented with the ``\texttt{io\_parallel}'' setting in hls4ml, where the full 500 samples are consumed at once.
For the MLP, we synthesize three reuse-factor settings in hls4ml: minimum reuse factor 1 (fully parallel, minimum II), mid-range reuse factor (II closer to that of the streaming SNNs for a fair comparison in resources), and maximum reuse factor (fully serialized, maximum II).
For the precision setting, the SNN models are quantized post-training and use fixed-point precision $\langle 16,6\rangle$ (16 total bits and 6 integer bits) for the weights and inputs, and $\langle 24,12\rangle$ for the membrane potentials; the MLP uses the bit widths learned during QAT.
The performance results presented in Sec.~\ref{sec:results} correspond to the floating-point SNN models and the QAT MLP, and the effect of the post-training quantization of the SNNs is discussed in Sec.~\ref{sec:results_separation}.
All models are synthesized with Vitis HLS (2023.2)~\cite{vitis} targeting the AMD Alveo U250 (FPGA part: ``xcu250-figd2104-2L-e'') at a clock frequency of 200 MHz (or 5 ns per clock cycle).
We discuss possible workarounds for the mismatch between the practical waveform sampling rate and the FPGA clock frequency in Sec.~\ref{sec:discussion}.
The FPGA resources are reported from the logic synthesis results with Vivado (2023.2).

\section{Results}
\label{sec:results}

\subsection{Model Prediction}
\label{sec:results_prediction}

\begin{figure}[!t]
    \centering
    \includegraphics[width=0.9\textwidth]{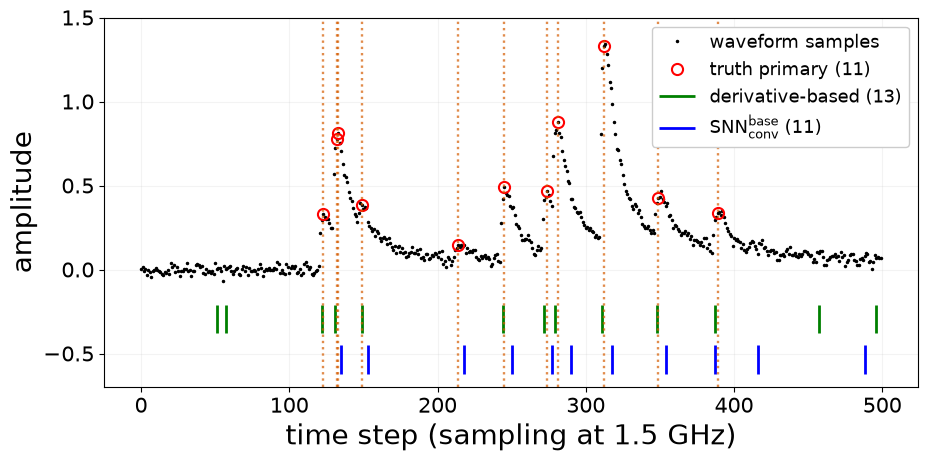}
    \caption{Output spikes predicted by $\text{SNN}_\text{conv}^{\text{base}}$ (blue) in a test waveform, where the truth primary clusters are shown (red). As a reference, the predicted primary clusters from the derivative-based method are also shown (green).}
    \label{fig:waveform_predict}
\end{figure}

We first examine if the SNN models are making sensible predictions by looking at the model outputs in a test waveform.
Fig.~\ref{fig:waveform_predict} shows the output spikes of $\text{SNN}_\text{conv}^{\text{base}}$ per time step in the waveform as well as the derivative-based predictions.
In this example waveform, there are 11 truth primary clusters, and $\text{SNN}_\text{conv}^{\text{base}}$ predicts the total number correctly.
Looking closely at the time steps when $\text{SNN}_\text{conv}^{\text{base}}$ fires, it can be seen that the output spikes show up several time steps after the truth primary clusters are consumed.
This is because of the internal signal integration in the LIF neurons: the membrane potential requires a few samples to accumulate and offset the leakage every time step to filter out fake peaks, and multiple layers add up the delay, so most output spikes come consistently some time steps after the peaks rather than promptly at the same time steps.

\begin{figure}[!t]
    \centering
    \includegraphics[width=0.6\textwidth]{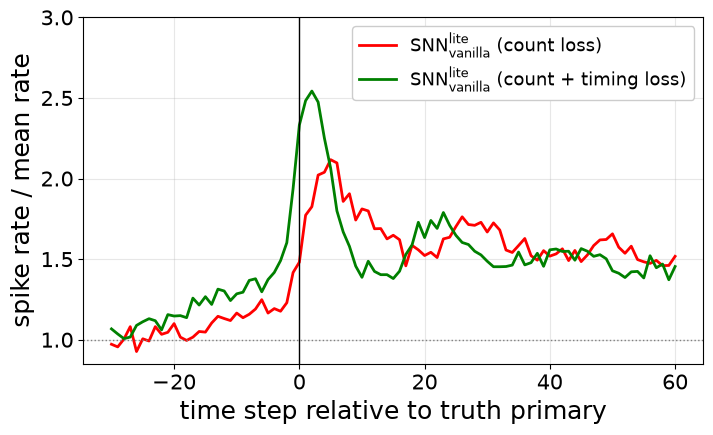}
    \caption{Output spike rate of $\text{SNN}_\text{vanilla}^{\text{lite}}$, normalized by the mean spike rate, as a function of the time steps relative to each truth primary cluster. The red curve corresponds to the model trained on the count only, and the green curve corresponds to the same model trained with an additional loss penalizing the output spikes landing in time steps far away from the truth primary clusters.}
    \label{fig:delay}
\end{figure}

Additionally, since the SNN models are trained with an MSE loss to encourage the number of output spikes to match the truth number without penalizing the timing, there is no explicit requirement that the SNN spikes occur exactly at the peak location.
As such, we validate that the models do not simply learn to regress the dataset mean but are indeed firing output spikes sensibly.
Fig.~\ref{fig:delay} shows, for $\text{SNN}_\text{vanilla}^{\text{lite}}$, the output spike rate divided by the mean spike rate as a function of the time step relative to each truth primary cluster, averaged over all primary clusters in 3000 test waveforms.
The model has a sharp rise in output spike rate well above the mean rate after the truth primary clusters arrive, demonstrating that the model does not simply regress the dataset mean, which would be indicated by a flat spike rate close to the mean rate.
We also compare with a reference model of the same architecture as $\text{SNN}_\text{vanilla}^{\text{lite}}$ trained with an additional loss that penalizes output spikes far away from the truth primary clusters.
It gives a sharper spike rate and a shorter delay, but the overall shape is consistent, validating the predictions of our SNN models.

\begin{figure}[!t]
    \centering
    \includegraphics[width=0.45\textwidth]{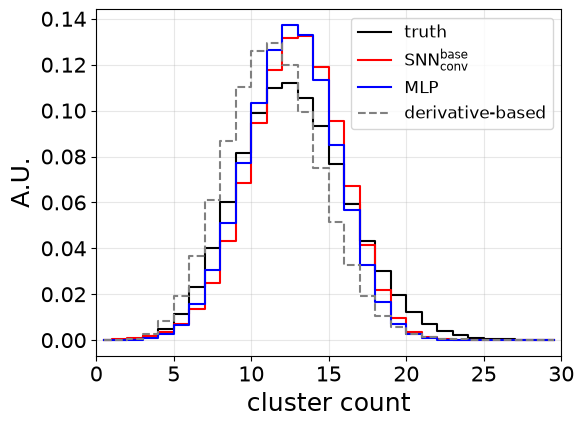}
    \caption{Predicted and truth cluster count distributions on the test set.}
    \label{fig:count_dist}
\end{figure}

\begin{figure}[!t]
    \centering
    \includegraphics[width=0.24\textwidth]{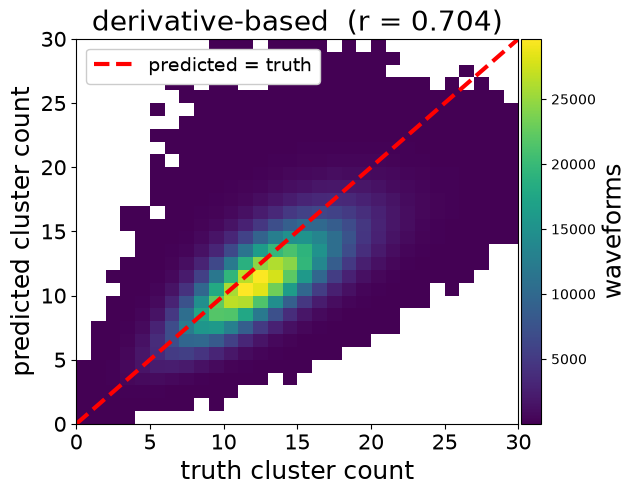}
    \includegraphics[width=0.24\textwidth]{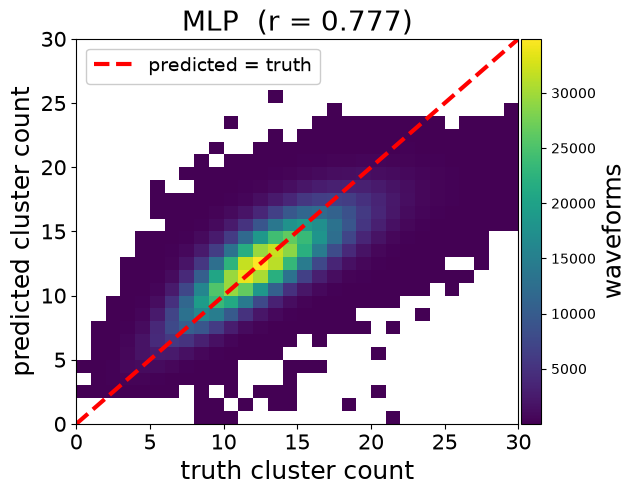}
    \includegraphics[width=0.24\textwidth]{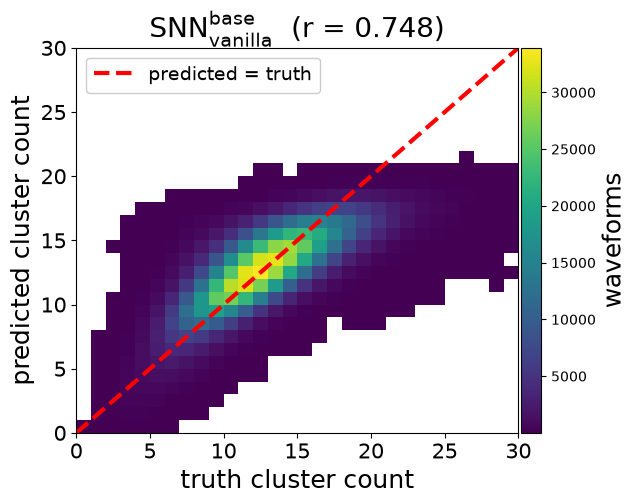}
    \includegraphics[width=0.24\textwidth]{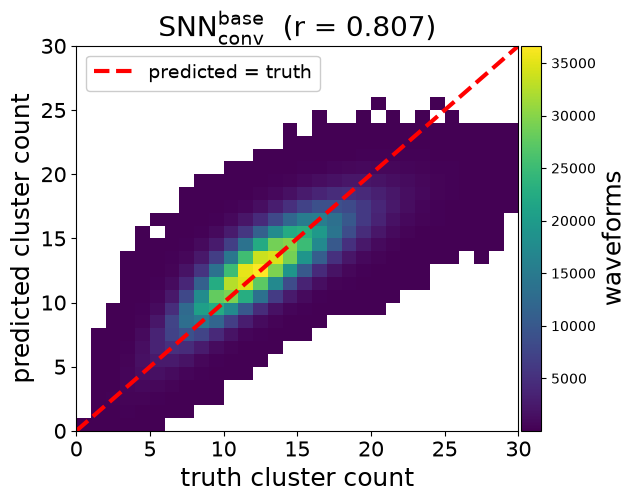}
    \caption{Predicted vs.\ truth cluster count on the test set and the correlation. From left to right: derivative-based (0.704), MLP (0.777), $\text{SNN}_\text{vanilla}^{\text{base}}$ (0.748), and $\text{SNN}_\text{conv}^{\text{base}}$ (0.807).}
    \label{fig:correlation}
\end{figure}

Fig.~\ref{fig:count_dist} shows the predicted and truth cluster count distributions for all test waveforms, where the $\text{SNN}_\text{conv}^{\text{base}}$ and MLP distributions are more compatible with the truth than that from the derivative-based method.
Fig.~\ref{fig:correlation} shows the correlation plots between the predicted cluster count and the truth cluster count.
All models show reasonable positive Pearson correlations.
The derivative-based method has the lowest correlation (0.704) and $\text{SNN}_\text{conv}^{\text{base}}$ has the highest (0.807), while the MLP is at 0.777 and $\text{SNN}_\text{vanilla}^{\text{base}}$ at 0.748.

\subsection{Separation Power}
\label{sec:results_separation}

From the predicted cluster count distributions of kaons and pions, we compute the separation power~\cite{Tian:2024yfz}:
\begin{equation}
    S=\frac{|\frac{dN}{dx}_{\text{kaon}}-\frac{dN}{dx}_{\text{pion}}|}{(\sigma_{\text{kaon}}+\sigma_{\text{pion}})/2},
\end{equation}
where $\frac{dN}{dx}$ and $\sigma$ are the mean number of primary ionization clusters per unit length and the standard deviation, respectively.
In our measurements, the cluster counts are obtained from a single drift cell, but one can scale the separation power to measurements over a longer track length.
When adding up the counts from multiple independent cells a track traverses, the total count scales linearly with the number of cells and thus with the track length.
Because the count follows Poisson statistics, the standard deviation scales with the square root of the mean count, and hence with the square root of the track length.
Combining these, the separation power scales as:
\begin{equation}
    S(L) = S_{\text{cell}} \sqrt{\frac{L}{L_{\text{cell}}}},
\end{equation}
where $S_{\text{cell}}$ is the separation power measured from a single cell, $L$ is the track length, and $L_{\text{cell}}$ is the mean track length per cell.
We take $L_{\text{cell}}=9.3$ mm for this dataset~\cite{Yilmaz:2025kqz}.

\begin{figure}[!t]
    \centering
    \includegraphics[width=0.6\textwidth]{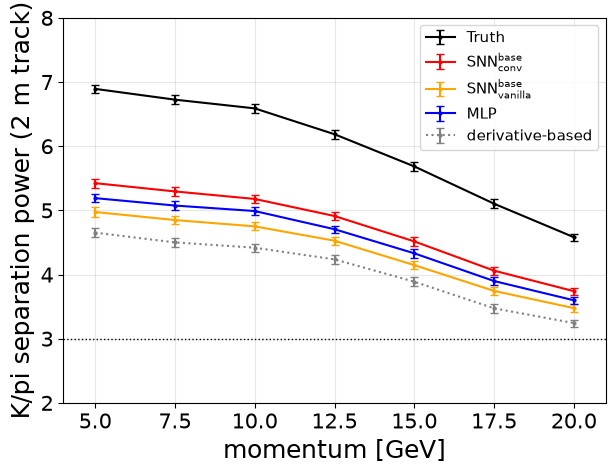}
    \caption{Kaon-pion separation power (scaled to a 2 m track length) as a function of momentum. The error bars correspond to the statistical uncertainty of one standard deviation. The $3\sigma$ requirement is shown.}
    \label{fig:separation}
\end{figure}

Fig.~\ref{fig:separation} shows the separation power, scaled to a 2 m track length~\cite{IDEAStudyGroup:2025gbt}, between kaons and pions in the seven testing momentum bins.
A minimum of 3$\sigma$ in kaon-pion separation power is required for CEPC and FCC-ee~\cite{FCC:2025lpp,CEPCStudyGroup:2025kmw}, and all models are above this floor across all momentum bins.
The performance ceiling is also obtained by computing $S$ using the truth information.
The ML models all outperform the derivative-based method, which has the lowest separation power in all momentum bins.
The $\text{SNN}_\text{conv}^{\text{base}}$ model yields the highest separation power in all momentum bins, even though its number of parameters is almost an order of magnitude smaller than that of the MLP baseline.
Tab.~\ref{tab:separation} lists the separation power of all models across the momentum bins, showing that all SNNs outperform the derivative-based method and that $\text{SNN}_\text{conv}^{\text{base}}$ has the best mean separation power, which is 17\% above the derivative-based method and 4\% above the MLP.
We also note that the MLP is trained with QAT, so its FPGA implementation is a bit-exact match to the trained model, whereas the SNNs are currently quantized after training, which results in a few percent drop in the mean separation power.
This gap is expected to be closed with QAT integration and is left for future work.

\begin{table*}[!t]
\caption{Kaon-pion separation power (2 m track length) of all models at each test momentum bin and the mean over all bins. Statistical uncertainty of one standard deviation is shown.}
\label{tab:separation}
\centering
\resizebox{\textwidth}{!}{
\begin{tabular}{llcccccccc} \toprule
    & \# param. & 5 GeV & 7.5 GeV & 10 GeV & 12.5 GeV & 15 GeV & 17.5 GeV & 20 GeV & mean \\ \midrule
    
    \textcolor{gray}{Truth} & $\textcolor{gray}{-}$ & $\textcolor{gray}{6.89 \pm 0.07}$ & $\textcolor{gray}{6.73 \pm 0.06}$ & $\textcolor{gray}{6.59 \pm 0.07}$ & $\textcolor{gray}{6.19 \pm 0.06}$ & $\textcolor{gray}{5.69 \pm 0.07}$ & $\textcolor{gray}{5.10 \pm 0.07}$ & $\textcolor{gray}{4.58 \pm 0.06}$ & $\textcolor{gray}{5.97 \pm 0.02}$ \\ \midrule
    
    $\text{SNN}_\text{vanilla}^{\text{lite}}$ & 49 & $4.88 \pm 0.07$ & $4.79 \pm 0.06$ & $4.66 \pm 0.07$ & $4.44 \pm 0.07$ & $4.10 \pm 0.06$ & $3.67 \pm 0.07$ & $3.42 \pm 0.07$ & $4.28 \pm 0.03$ \\
    
    $\text{SNN}_\text{vanilla}^{\text{base}}$ & 321 & $4.97 \pm 0.07$ & $4.85 \pm 0.06$ & $4.75 \pm 0.06$ & $4.53 \pm 0.07$ & $4.15 \pm 0.07$ & $3.74 \pm 0.07$ & $3.48 \pm 0.06$ & $4.35 \pm 0.02$ \\
    
    $\text{SNN}_\text{conv}^{\text{lite}}$ & 73 & $5.03 \pm 0.07$ & $4.92 \pm 0.06$ & $4.83 \pm 0.07$ & $4.55 \pm 0.07$ & $4.20 \pm 0.06$ & $3.81 \pm 0.07$ & $3.53 \pm 0.06$ & $4.41 \pm 0.03$ \\
    
    $\text{SNN}_\text{conv}^{\text{base}}$ & 529 & $\textbf{5.42} \pm \textbf{0.07}$ & $\textbf{5.30} \pm \textbf{0.06}$ & $\textbf{5.18} \pm \textbf{0.06}$ & $\textbf{4.91} \pm \textbf{0.07}$ & $\textbf{4.52} \pm \textbf{0.07}$ & $\textbf{4.06} \pm \textbf{0.07}$ & $\textbf{3.74} \pm \textbf{0.06}$ & $\textbf{4.73} \pm \textbf{0.02}$ \\ \midrule
    
    MLP & 4569 & $5.19 \pm 0.07$ & $5.08 \pm 0.07$ & $4.99 \pm 0.07$ & $4.71 \pm 0.07$ & $4.33 \pm 0.07$ & $3.90 \pm 0.07$ & $3.60 \pm 0.07$ & $4.54 \pm 0.02$ \\
    
    derivative-based & $-$ & $4.65 \pm 0.07$ & $4.50 \pm 0.07$ & $4.42 \pm 0.06$ & $4.24 \pm 0.06$ & $3.89 \pm 0.07$ & $3.47 \pm 0.07$ & $3.24 \pm 0.06$ & $4.06 \pm 0.02$ \\ \bottomrule
    
\end{tabular}
}
\end{table*}

\subsection{FPGA Synthesis Results}
\label{sec:results_fpga}

\begin{figure}[!t]
    \centering
    \includegraphics[width=0.8\textwidth]{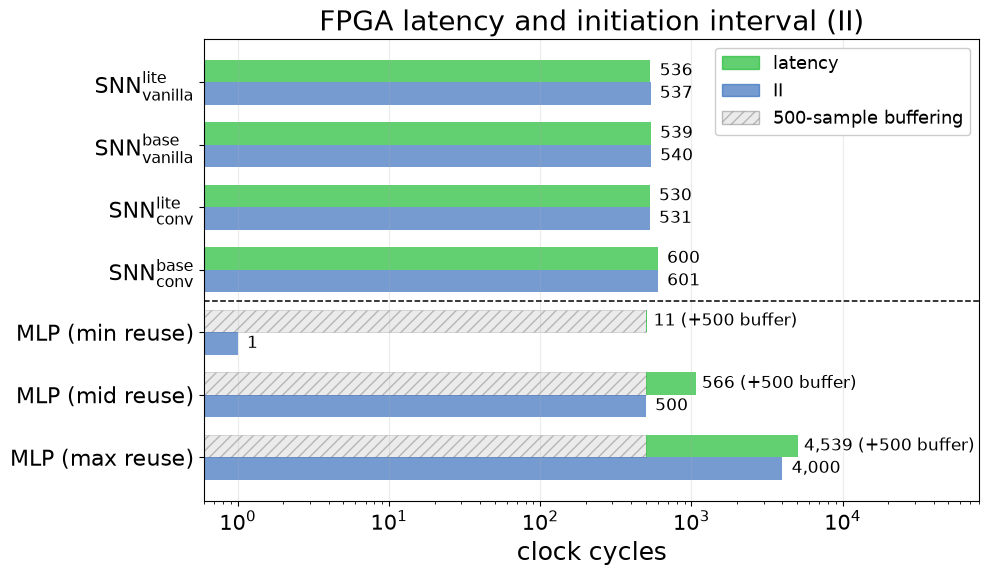}
    \caption{Latency and II in clock cycles (200 MHz, or 5 ns) on an FPGA. The MLP must receive the full waveform before processing and so has a 500-sample buffer added, which corresponds to an additional 500 clock cycles (hatched) assuming one sample arrives per clock cycle. The streaming SNNs can process the samples upon arrival.}
    \label{fig:latency}
\end{figure}

We present the FPGA synthesis results and show that the SNN models use significantly fewer resources than the MLP baseline at a comparable II.
Fig.~\ref{fig:latency} shows the model latency for processing one waveform on an FPGA.
Because the SNN models are designed to process one sample per time step, and the waveform samples are assumed to arrive one per clock cycle, the streaming SNNs process each sample as soon as it arrives without the need for extra buffering.
The dense layers of the MLP require the full input before they can propagate, so the MLP stays idle on-chip until all 500 samples have arrived, and this buffering window of 500 clock cycles is added to its latency in Fig.~\ref{fig:latency}.
Note that the fully parallelized MLP (minimum reuse factor) has an II of one clock cycle.
This only means that the model itself is fully pipelined, and in practice it still needs to wait for the full waveform to arrive before processing.
As a comparison, $\text{SNN}_\text{vanilla}^{\text{base}}$ and $\text{SNN}_\text{conv}^{\text{base}}$ require 539 and 600 clock cycles to process a 500-sample waveform, while the MLP at full parallelization requires 500+11 clock cycles, and as the reuse factor increases, the latency increases dramatically as expected.
Overall, the streaming SNN models return the cluster count 30 to 100 clock cycles after the last sample of the waveform is consumed, which is 150 to 500 ns at a 5 ns clock, compatible with a continuous readout system that handles its own data buffering.

\begin{figure}[!t]
    \centering
    \includegraphics[width=\textwidth]{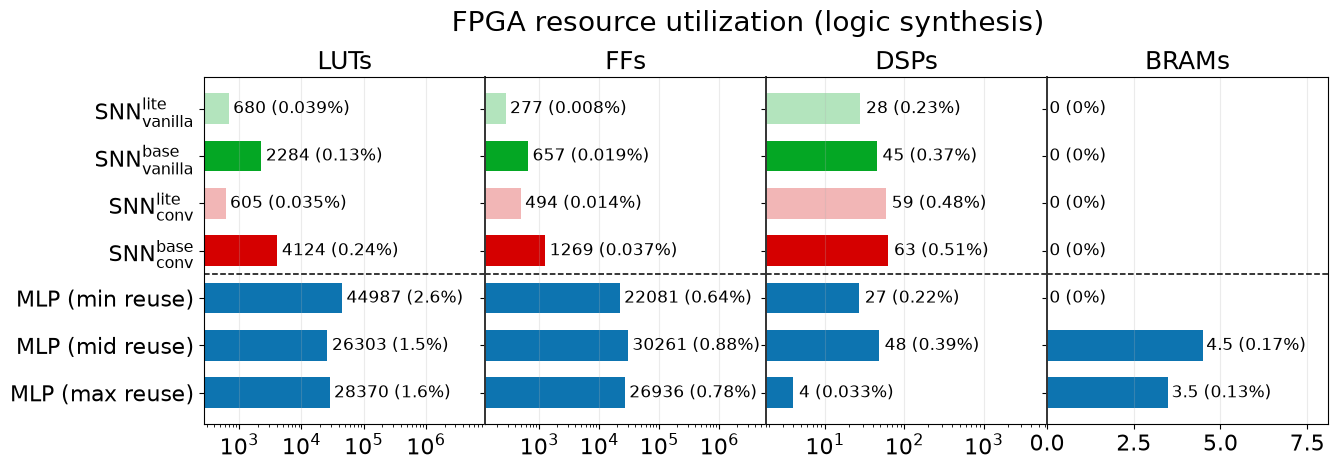}
    \caption{FPGA resource utilization (after logic synthesis). The numbers in parentheses refer to the utilization over the total available on-chip resources.}
    \label{fig:resources}
\end{figure}

Fig.~\ref{fig:resources} shows the post-synthesis FPGA resource estimates for the models in terms of lookup tables (LUTs), flip-flops (FFs), digital signal processing blocks (DSPs), and block RAMs (BRAMs).
All SNN models use roughly an order of magnitude fewer LUTs and FFs than the MLP: e.g.\ $\text{SNN}_\text{conv}^{\text{base}}$ uses 4124 LUTs and 1269 FFs, while the fully parallelized MLP uses 44987 LUTs and 22081 FFs, and the MLP (mid reuse factor) with an II comparable to the SNNs uses 26303 LUTs and 30261 FFs.
This is because the spikes are binary, so a dense or convolutional layer following a spiking neuron layer becomes multiply-free.
Also, only a small number of membrane potentials are kept in local registers and a full waveform buffer is not needed, so no BRAM is used.
The MLP uses few DSPs because QAT compresses its precision to small bit widths, whereas the SNN models use their DSPs mainly for the high-precision input-weight multiplications in the first layer and for the membrane potential updates.
Regardless, the DSP utilization for the SNN models is at most around 0.5\% of the total available on-chip DSPs, which could likely be further reduced if the SNNs were trained with QAT.

\section{Discussion}
\label{sec:discussion}

\textbf{SNNs excel at PID even at tiny model sizes.}
Previous approaches to cluster counting either use models that are too large to meet real-time processing constraints or trade PID performance for a smaller size that fits on an FPGA.
Our streaming SNN architectures show that such a trade-off is not necessary.
With only 49 to 529 parameters, all SNN models achieve a mean kaon-pion separation power well above the $3\sigma$ requirement.
In particular, $\text{SNN}_\text{conv}^{\text{base}}$ with 529 parameters gives the highest separation power of all models: 17\% higher than the derivative-based method and 4\% higher than an MLP that has $8.6\times$ more parameters.
This suggests that SNNs are well-suited to the cluster counting task, with better PID performance than existing methods.

\textbf{SNNs fit well on FPGAs for real-time applications.}
What makes SNNs efficient on hardware is the sparsity of their signal representation and processing.
The spiking neurons output spikes that are 1-bit, so the spike-weight products of a dense or convolutional layer following a spiking neuron layer are simplified to conditional additions (multiply-free).
While not done in this work, one could also fix the leak factor in LIF neurons to be of the form $1-2^{-k}$, so that the membrane leak operation simplifies to a bit shift and a subtraction in hardware.
The streaming formulation we have developed also allows the SNNs to process each sample as soon as it arrives, avoiding the need for buffering a full waveform.
Additionally, the SNN has far fewer weights and applies the same set of weights to every sample, emitting binary spikes between layers and at the model output.
This lightweight design is why the SNNs use an order of magnitude fewer LUTs and FFs than the MLP.
We note that the same streaming approach could also potentially be implemented with other architectures such as the MLP we presented, but even in this case it would lack the sparse computation advantage of the SNNs.

\textbf{$O(200\,\text{MHz})$ FPGA clock frequency vs.\ $O(\text{GHz})$ waveform sampling rate.}
The simulated dataset sets the sampling rate at 1.5 GHz, which is required to resolve neighboring cluster peaks that are separated within a few ns.
However, such a sampling rate already exceeds the practical clock frequency one can run on modern FPGAs.
In the experiments presented here, the 1.5 GHz sampling rate is $7.5\times$ the 200 MHz clock we configure for synthesis, where we naively assumed the samples were streaming in at the same 200 MHz rate.
This can possibly be solved by designing a firmware block that consumes and processes several consecutive samples instead of one per clock cycle, at the cost of more resources to handle the loop-carried dependency of the membrane potential updates across time steps, and of a higher latency if a lower clock frequency is required.
These are likely affordable given that the current design is already very low in resources and latency.
Multiple copies of the SNN could also be employed in a parallel processing scheme, allowing the design to keep up with the full data rate, again at the cost of resources for these copies and the additional routing.
Alternatively, the $\text{SNN}_\text{conv}$ architecture could also be modified to use a stride of 8 (with a kernel size of at least 8 as well) in the first layer, thereby allowing the architecture to keep up with the full data rate by processing the arrived samples within one 200 MHz clock cycle.
Such designs would need to be implemented and tested.
In addition, one could consider implementations on an ASIC with a higher clock frequency.

\textbf{SNNs could also be implemented in analog hardware for ultra-low power.}
To further unleash the SNN advantages, one can consider neuromorphic chips with analog circuits on which spiking neurons are directly embedded.
This could eliminate the need for a digitization step that runs at an $O(\text{GHz})$ sampling rate and requires significant power to produce digital signals that need to be transmitted through optical links to a separate FPGA/ASIC board on which the cluster counting algorithm is run.

\section{Summary}
\label{sec:summary}

At a time when future collider proposals are being actively studied, the next generation of drift chambers is being designed to enable the cluster counting technique for $dN/dx$-based PID.
However, the high sampling rates needed to resolve the individual ionization clusters for this method may require on-detector reconstruction to reduce the readout bandwidth. 
We propose SpikePID, a neuromorphic approach using spiking neural networks (SNNs) to perform real-time cluster counting in future drift chambers. 
We have developed a streaming HLS implementation of SNNs built upon hls4ml, so that an end-to-end workflow from model training to FPGA deployment is available for further study and real applications.
This work provides a proof of concept that a streaming SNN architecture with a few hundred parameters already yields better kaon-pion separation power than the traditional derivative-based method and an MLP with thousands of parameters, while using an order of magnitude fewer resources on an FPGA.
The next steps include a systematic study of the effects of different drift chamber configurations such as variations in the gas mixture and the drift cell geometry, a validation of the models on waveforms from test beams with realistic background noise, and the actual implementation of these models in real digital or analog readout systems.

\section*{Acknowledgments}
HFT and DSR are supported by the U.S. Department of Energy (DOE), Office of Science, Office of High Energy Physics subprogram on Computational High Energy Physics under Award No. DE-SC0026801.
DSR is also supported by the U.S. Department of Energy (DOE), Office of Science, Office of High Energy Physics Early Career Research program under Award No. DE-SC0025324.
This work used resources available through the National Research Platform (NRP) at the University of California, San Diego~\cite{10.1145/3708035.3736060}.
NRP has been developed, and is supported in part, by funding from National Science Foundation, from awards 1730158, 1540112, 1541349, 1826967, 2112167, 2100237, and 2120019, as well as additional funding from community partners.

\bibliographystyle{naturemag}
\bibliography{references}


\end{document}